\documentclass[epj,nopacs]{svjour}
\usepackage{float}
\usepackage{amsmath,amssymb}
\usepackage{graphicx}
\usepackage{xcolor}
\usepackage[utf8]{inputenc}
\DeclareUnicodeCharacter{2212}{-}   
\usepackage{amsmath,amssymb}
\usepackage[colorlinks=true, linkcolor=blue, citecolor=blue, urlcolor=blue]{hyperref}   
\usepackage[T1]{fontenc}      
\usepackage{textcomp}      
\usepackage[numbers,sort&compress]{natbib}
\usepackage{doi}
\usepackage{booktabs}
\usepackage{multirow}

\begin{document}
\sloppy
\title{Reconstructing Gamma-Ray Burst Energy Relations with Observational $H(z)$ Data in a Neural Network Framework}

\author{
Nilanjana Bagchi Aurpa\inst{1} \and
Abha Dev Habib\inst{1} \and
Nisha Rani\inst{1}
}

\institute{
Miranda House, University of Delhi, Delhi 110007, India \\
\email{\url{abhadev.habib@mirandahouse.ac.in}
}}
\titlerunning{GRB Energy Relations with OHD in NN Framework}
\authorrunning{Aurpa N.B,Habib A.D, Rani N.}
\date{Received: date / Revised version: date}

\abstract{Gamma-ray bursts (GRBs) offer a powerful probe of the cosmic expansion history far beyond the redshift range accessible to Type Ia supernovae. However, the study of cosmological models using GRBs is hindered by the circularity problem, which arises from assuming a fiducial cosmological model during GRB luminosity distance calibration. In this work, we perform a model-independent calibration of GRB luminosity relations using observational measurements of the Hubble parameter from the A220 and J220 compilations, thereby avoiding explicit cosmological assumptions. We employ an Artificial Neural Network to reconstruct the calibration relation directly from the data. In addition, we implement a Bayesian Neural Network framework as an alternative approach, enabling a data-driven treatment of both statistical and systematic uncertainties. We use calibrated GRB samples to constrain the Amati relation and systematically compare the outcomes obtained from different calibration techniques and datasets. We find that the Amati relation slopes derived from the two neural network approaches are consistent with each other and with previous low-redshift calibrations obtained using model-independent methods. The Bayesian Neural Network approach provides a more robust framework for propagating uncertainties in the calibration procedure.} 
\maketitle

\section{Introduction}
Gamma-ray bursts (GRBs) are among the most energetic high-energy events in the Universe and can be detected at extremely large cosmological distances, with confirmed observations extending up to redshifts of $z \sim 9.4$ \cite{Salvaterra2009,Cucchiara2011}. 
This redshift range significantly exceeds that of Type Ia supernovae (SNe Ia), which currently populate the Hubble diagram only up to $z \sim 2$ \cite{Scolnic2018,Scolnic2022}. 
Consequently, GRBs provide a unique opportunity to extend cosmological distance measurements to much earlier cosmic probes.\\

The use of GRBs for cosmological applications is based on several empirical energy-luminosity correlations, such as the Amati, Ghirlanda, Yonetoku,Liang-Zhang Correlations\cite{Amati2002,Ghirlanda:2004fs,Yonetoku:2003gi,Izzo:2015vya,Liang2005} and Plateau based correlations - Dainotti Relations\citep{Dainotti2008,Dainotti2011,
Dainotti2013,Dainotti2016,Dainotti2018,
Dainotti2021}. The Amati relation \cite{Amati2002,Amati2008} is the correlation between the peak and isotropic energy ($E_{\mathrm{p}}$--$E_{\mathrm{iso}}$) of the GRBs. The isotropic energy of a GRB is a derived quantity which depends on the luminosity distance and observed bolometric flux. As luminosity distance is model dependent, early studies typically calibrated these correlations by assuming a fiducial cosmological model, largely the $\mathrm{\Lambda}$CDM model. GRB correlations calibrated by assuming an underlying cosmological model is then used to constrain parameters of cosmological models.  This approach leads to the circularity problem.\\

Several methods have been proposed to overcome this limitation. To address the circularity problem, Liang et al. (2008) \cite{Liang2008} introduced a model-independent calibration approach in which GRB distances are obtained by interpolating from low-redshift SNe Ia observations, without assuming a specific background cosmology. 
An alternative strategy is the simultaneous fitting method \cite{Amati2008,Wang2008}, where the parameters of the GRB luminosity relations and the cosmological model are constrained jointly within a single statistical framework. 
Since the resulting GRB correlation parameters do not exhibit strong dependency to the underlying cosmological assumptions, these studies indicate that GRBs can be reliably standardized within current observational uncertainties \cite{Khadka2020}.\\

Independent observational datasets have also been used for GRB calibration. 
Amati et al. \cite{Amati2019} employed Observational Hubble Data (OHD) derived from the Cosmic Chronometer (CC) method and used a Bezier parametric reconstruction to calibrate the Amati Relation \cite{Amati2019,Amati2008}. The calibrated GRB relation has subsequently been employed in a number of studies to place constraints on cosmological models using independent observational data \citep{Montiel:2020rnd,Luongo:2021pjs,Luongo:2022bju,Muccino:2020gqt,Muccino:2022rnd}.\\

In parallel, a broad spectrum of calibration methodologies have been developed. These include interpolation-based techniques \cite{Liang2008,Liu:2022inf}, local regression schemes \citep{Cardone:2009mr,Demianski:2016zxi,Demianski:2019vzl}, Bezier parametric reconstructions \citep{Amati2019}, iterative calibration procedures \citep{Liang2008}, and approaches based on Pad\'e approximations \cite{Liu:2014vda}. 
Within the class of non-parametric methods, Gaussian Process regression has emerged as a widely used tool for model-independent cosmological reconstruction and represents one of the earliest applications of machine learning techniques in this field \cite{Seikel2012a,Seikel:2012cs,Li:2017zrx,Han2024,Pan:2020zbl,Mu:2023bsf,Mu:2023zct,Sun:2021pbu,Zhang:2024ndc,Kumar:2022ypo}. 
Despite its flexibility, Gaussian Process regression is sensitive to the kernel choice, which may impact the reliability of the reconstructed functions \citep{Zhang:2023xgr,Wei:2016xti, Johnson2025blf}.\\

In recent studies, Artificial Neural Networks (ANNs) have been proposed as an alternative framework for cosmological reconstruction \cite{Luongo:2020hyk}. 
Compared to Gaussian Processes which assumes Gaussian Distribution implicitly and is influenced by the choice of kernel, ANNs are inherently more data-driven and impose significantly fewer assumptions on the properties of the data, which has motivated their increasing use in cosmological applications \cite{Zhang:2024ndc,Dialektopoulos:2021wde}. This allows for the reconstruction of functions directly from observational data without assuming an explicit functional form as demonstrated by Wang et al. (2019) \cite{Wang2019} who reconstructed the Hubble parameter as a function of redshift using OHD.\\

ANN architecture and training process consist of  many hyperparameters. This poses a challenge to minimise the loss. Furthermore, ANN do not inherently provide the uncertainty associated with the predictions and it may confidently predict inaccurate values. This may make it unreliable for observational cosmological datasets which generally have large uncertainty like Observational Hubble Dataset \cite{Mahida:2025teg} . \\

Bayesian formulations of neural networks, originally developed by Bishop et al. \citep{bishop2006prml,bishop2013brml}, provides a principled framework for propagating uncertainty from model parameters to predicted observable \citep{Gelman2013,ghahramani2015probabilistic}. The Bayesian evidence naturally encodes Occam’s razor by penalising overly flexible or excessively complex models and reduces the model bias \cite{Neal1996}. Further,   BNN employs Bayesian Inference to find the optimum parameters for the datasets by learning the distribution of the network parameters instead of point values. Thus providing both the prediction and the related uncertainty estimation.\\

In this work, we calibrate GRBs using OHD by employing Artificial Neural Networks, building on the approaches adopted in previous studies \cite{Wang2019,Huang2025h,Shah2024}. Bootstrap sampling is introduced to improve the estimation of uncertainties in the predicted values. As a complementary approach to quantify model uncertainty, we also employ a Bayesian Neural Network (BNN). Unlike standard ANN implementations, BNNs treat the network parameters probabilistically and naturally propagate uncertainties through the posterior distribution. The simultaneous application of ANN and BNN frameworks, therefore,  provides an additional consistency test for the reconstructed Hubble parameter and allows us to assess the robustness of the resulting GRB calibration against the choice of machine-learning methodology. We construct the most suitable ANN model that are capable of reproducing the observed Hubble parameter data using a grid-search using RISK function. For BNN model, we utilise  Widely Applicable Information Criterion (WAIC) as criterion for BNN which provides a principled balance between predictive accuracy and model complexity. Following the reconstruction, we calibrate the GRBs using both ANN and BNN predictions. The resulting calibrations are then used to estimate the Amati relation parameters through Markov Chain Monte Carlo analysis, allowing a comparison of the efficiency and reliability of the two machine-learning approaches in GRB calibration.\\

This paper is structured as follows. We first calibrate GRBs using OHD with an Artificial Neural Network, following methodologies established in earlier studies as described in Section~\ref{subsec:ann}. In Section~\ref{subsec:bnn}, we present an alternative calibration based on a Bayesian Neural Network framework. Finally, in Section~\ref{subsec:amati}, the calibrated GRB sample is used to constrain the Amati relation. Results obtained from different calibration methods and datasets are shared in Section \ref{sec:results}. We end the paper with conclusion and discussion in Section \ref{conclusion}.

\section{Data and Methodology}
\subsection{Data}
For our study to calibrate Gamma-Ray Bursts, we utilise the updated Hubble data from Table 1 of Ratra et al. \cite{Ratra2023}. The dataset comprises 32 data
points within a redshift range of $0.07 < z < 1.965$.\\

The Amati relation is applicable only to long GRBs, defined by a rest-frame duration $T_{90,\mathrm{rest}} > 2~\mathrm{s}$. In this analysis, we chose two GRB datasets individually to check the robustness of our approach. For the first set, we consider A220 datset consisting of 220 long GRBs. This dataset is a combination of two earlier subsets, i.e. A118 and A102, from Table 7 and 8 respectively from Khadka et al \cite{Khadka:2021vqa}. The A220 dataset has been considered standard for cosmological analysis purposes and have been utilised widely in previous studies \citep{Kumar:2022ypo,Huang2025h, Han2024}.\\

The second data, referred to as J220 dataset is taken from Jia et al. \cite{Jia2022}. It comprises recent GRBs from Swift\footnote{\url{https://swift.gsfc.nasa.gov/archive/grb_table.html}} and Fermi\footnote{\url{https://heasarc.gsfc.nasa.gov/FTP/fermi/data/gbm/daily/}} catalog, alongside previous subsamples. Though the A220 and J220 samples partially overlap but they differ in their construction, data sources, and selection criteria. Using both allows us to assess the calibrated GRB correlations against catalog-dependent systematics. \\

Due to the use of low redshift cosmological datasets, which only go up to $z\sim 1.965$. We consider GRBs with redshift below 1.965 to be used for calibration purpose and put constraint on the Amati Relation parameters. We utilise 115 GRBs from A220 dataset and  129 GRBs from J220 dataset. 

\subsection{Reconstruction of H(z) with Artificial Neural Network}
\label{subsec:ann}
An Artificial Neural Network is built from interconnected neurons arranged in input, hidden, and output layers. The Input layer consists of the features per datapoint we want to analyse. The Hidden layer section is where the features of NN (Neural Network) come as it allows to capture complex relations in data. An ANN can have multiple layers depending on data complexity. For the first hidden layer, the output of each neuron depends on the input datapoints and is constructed as a linear function of the inputs with a weight and bias associated with the neuron.\\

In ANN, the output of each neuron is defined by a set of weights and biases, where the output of a neuron in one layer is dependent on the output of all neurons in the previous layer. The linear combination of the outputs of the neurons is passed through an activation function which introduces non-linearity to the relation of weights.\\

The relation is given by:
\begin{equation}
\begin{aligned}
\text{Layer input: } & z^{(l)} = W^{(l)} a^{(l-1)} + b^{(l)} \text{ and } \\
\text{Layer output: } & a^{(l)} = f\big(z^{(l)}\big).
\end{aligned}
\end{equation} Here, f(z) is the activation function and W and b represent the weights and the biases. \\

The output section provides the required features and completes the forward propagation section of ANN. After the forward propagation, it undergoes a training process to parametrise the weights that are initially chosen at random. 
A back-propagation algorithm is used, where a suitable loss function is selected, and the weights are updated using optimizers such as ADAM, SGD, or Gradient Descent.
The weight update is:
\begin{equation}
\theta \leftarrow \theta - \eta\,\nabla_\theta \mathcal{L},
\end{equation}
where $\eta$ is the learning rate which can be fixed or dynamic. It can be adjusted during training by using schedulers  during training  depending on multiple parameters, including the number of iterations, and behavior of the loss, for example exponential decay.\\

\begin{figure}[t]
    \centering
    \resizebox{0.40\textwidth}{!}{
    \includegraphics[width=0.5\textwidth]{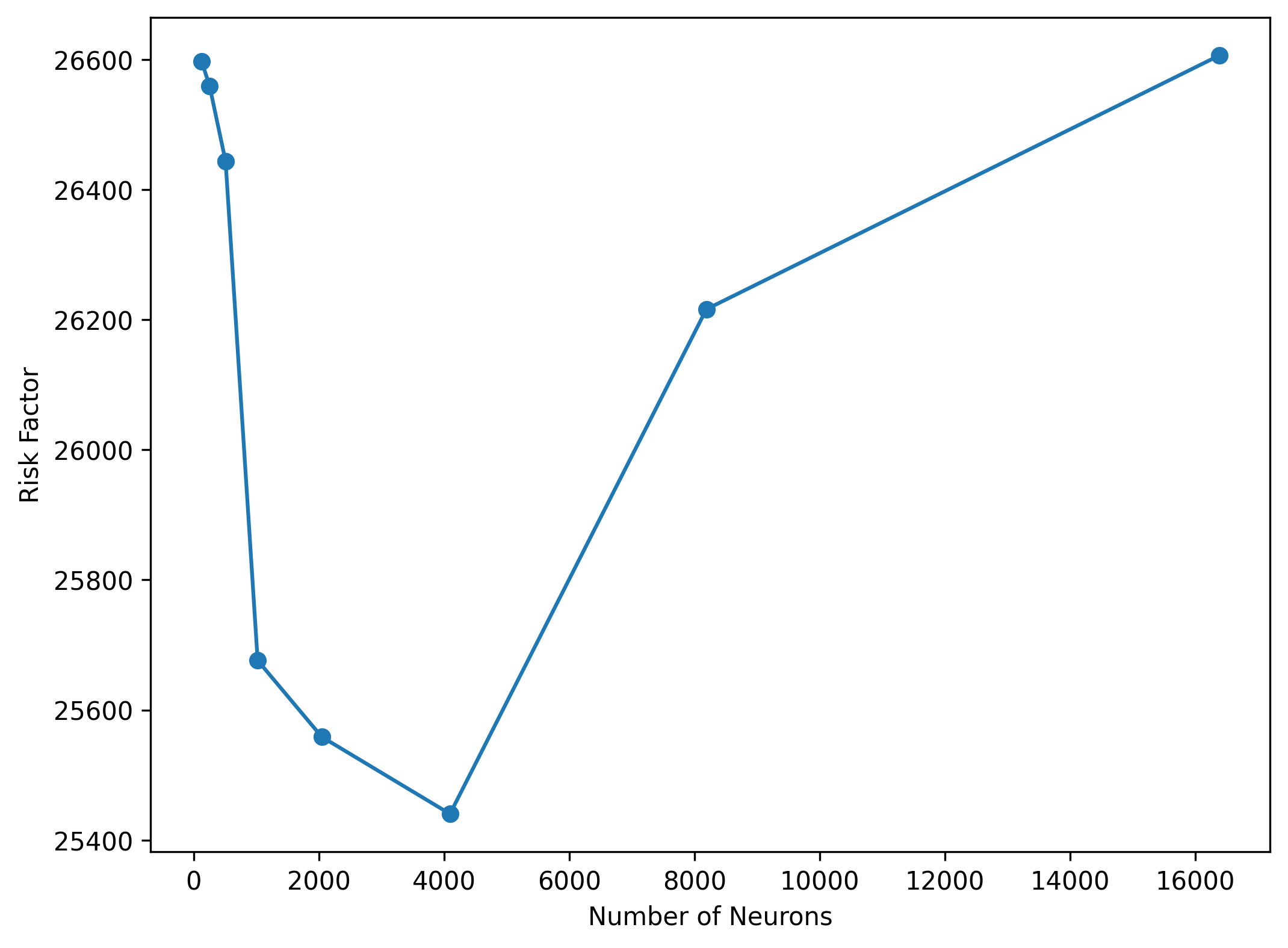}
}
   \caption{Risk statistic as a function of the number of neurons in the hidden layer of the ANN architecture.The minimum of the curve, 4096, determines the optimal network configuration used for the Hubble parameter reconstruction.}
    \label{fig:neuron}
\end{figure}

To determine the optimal hyperparameters of the ANN model, grid search is performed using the RISK function as formulated in statistical decision theory \cite{Wasserman2001ng}. In this framework, the risk corresponds to the expected squared prediction error and is commonly used for regression problems with Gaussian uncertainties. The adopted loss function is therefore given by
\begin{equation}
\text{RISK} = \sum_i \left[ (H_{p,i} - H_{z,i})^{2} + \sigma_{z,i}^{2} \right],
\end{equation}
where $H_{p,i}$ represents the predicted value and $H_{z,i}$ the observed Hubble parameter with uncertainty $\sigma_{z,i}$. Minimising this quantity corresponds to selecting the model that minimises the expected prediction error while accounting for observational uncertainties.Unlike previous works that employ simulated datapoints for hyperparameter optimisation \cite{Wang2019,Huang2025h}, we use the full OHD dataset as the validation set, allowing the grid search to be directly guided by the empirical data distribution.\\

The numbers of neurons is evaluated for a single hidden layer varying as \(2^n\) for \(7 \le n \le 14\). The results of RISK values for different neuron counts are shown in Figure \ref{fig:neuron}. The optimal neuron count is found to be \textbf{4096}, consistent with earlier works by \cite{Huang2025h,Wang2019} and other Hubble-parameter reconstruction studies \cite{Wei:2022plg}.\\

\begin{figure*}

    \centering
    \begin{minipage}[b]{0.48\textwidth}
        \centering
        \includegraphics[width=\textwidth]{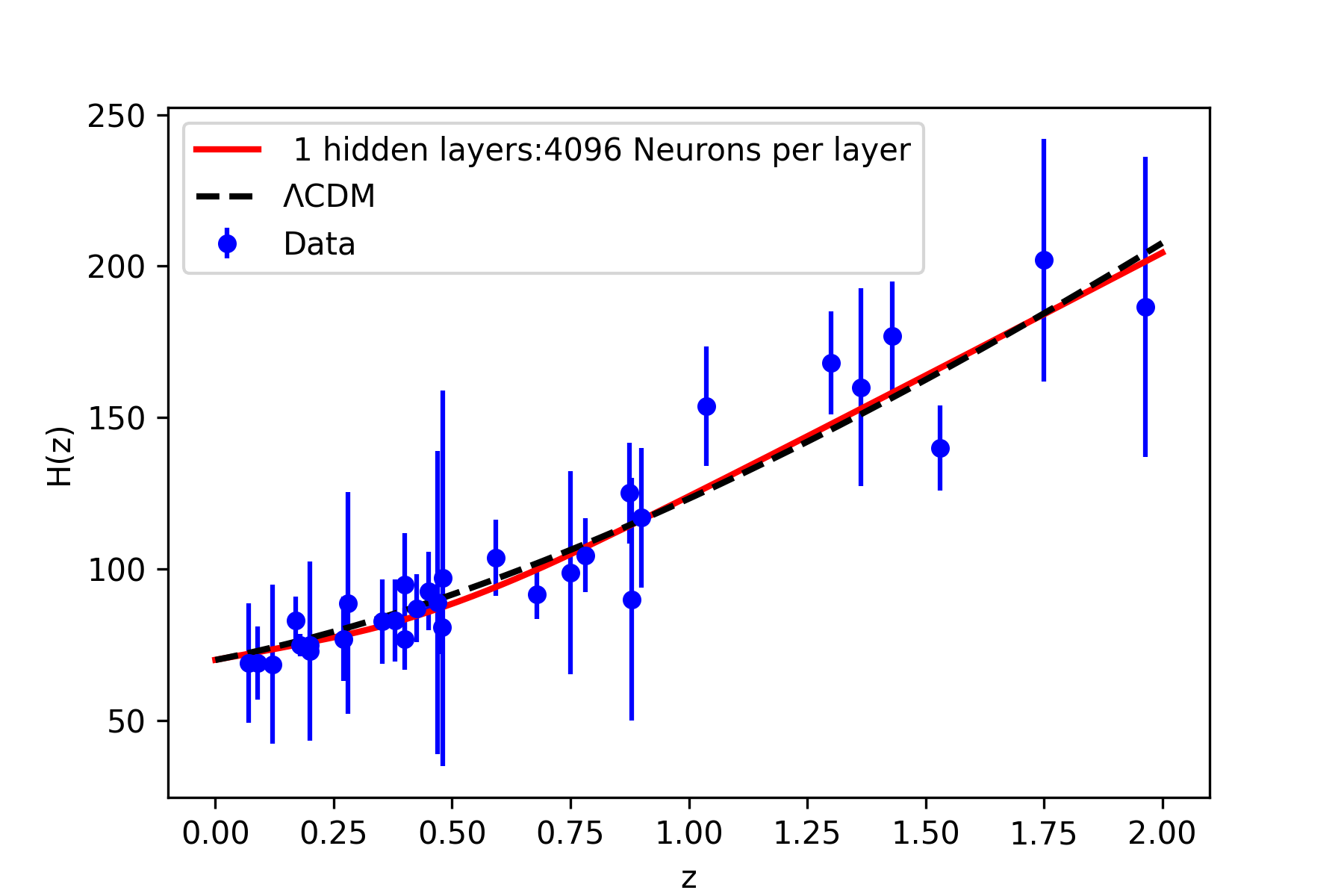}
        (a)
    \end{minipage}
    \hfill
    \begin{minipage}[b]{0.48\textwidth}
        \centering
        \includegraphics[width=\textwidth]{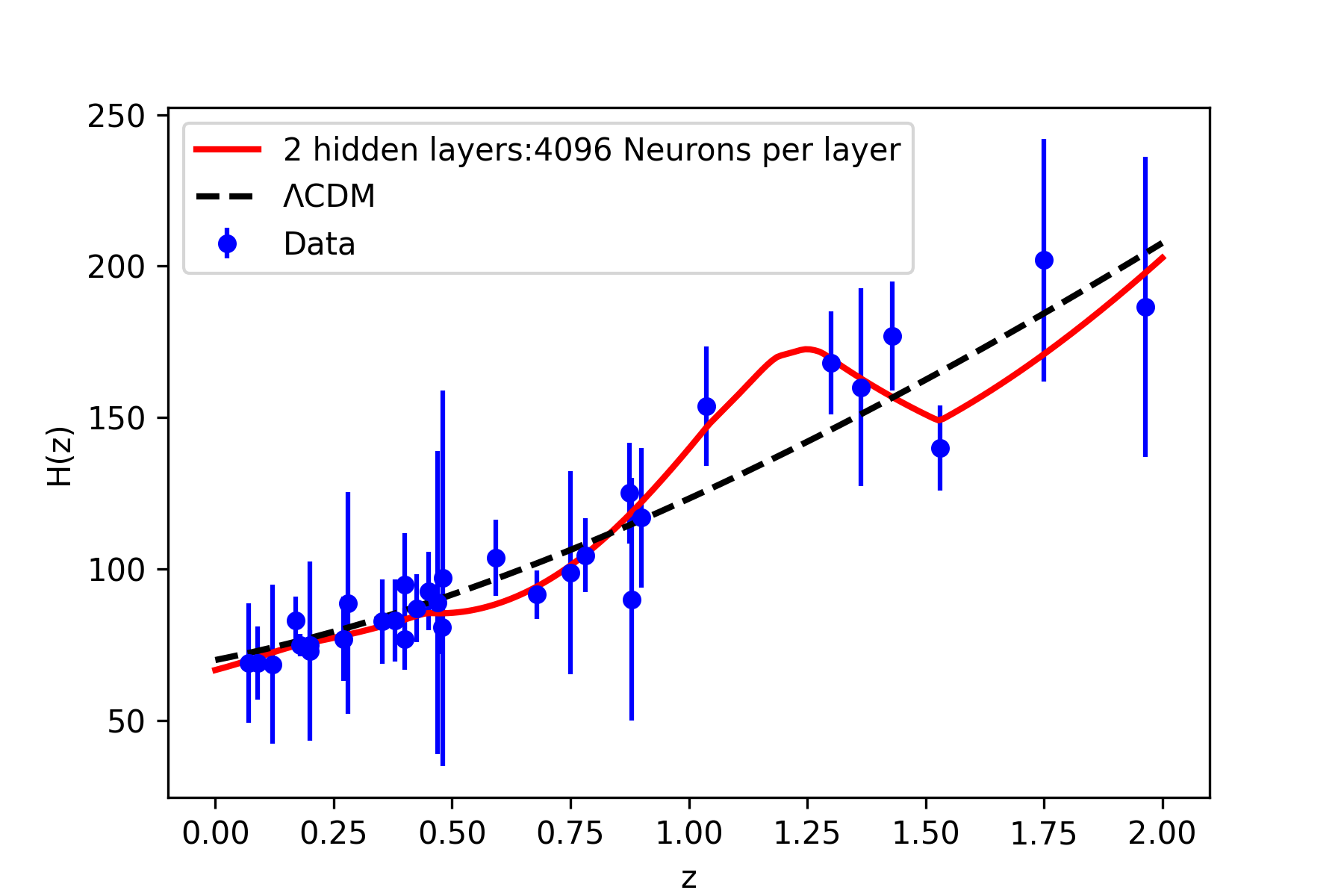}
        (b)
    \end{minipage}
    \hfill
    \begin{minipage}[b]{0.48\textwidth}
        \centering
        \includegraphics[width=\textwidth]{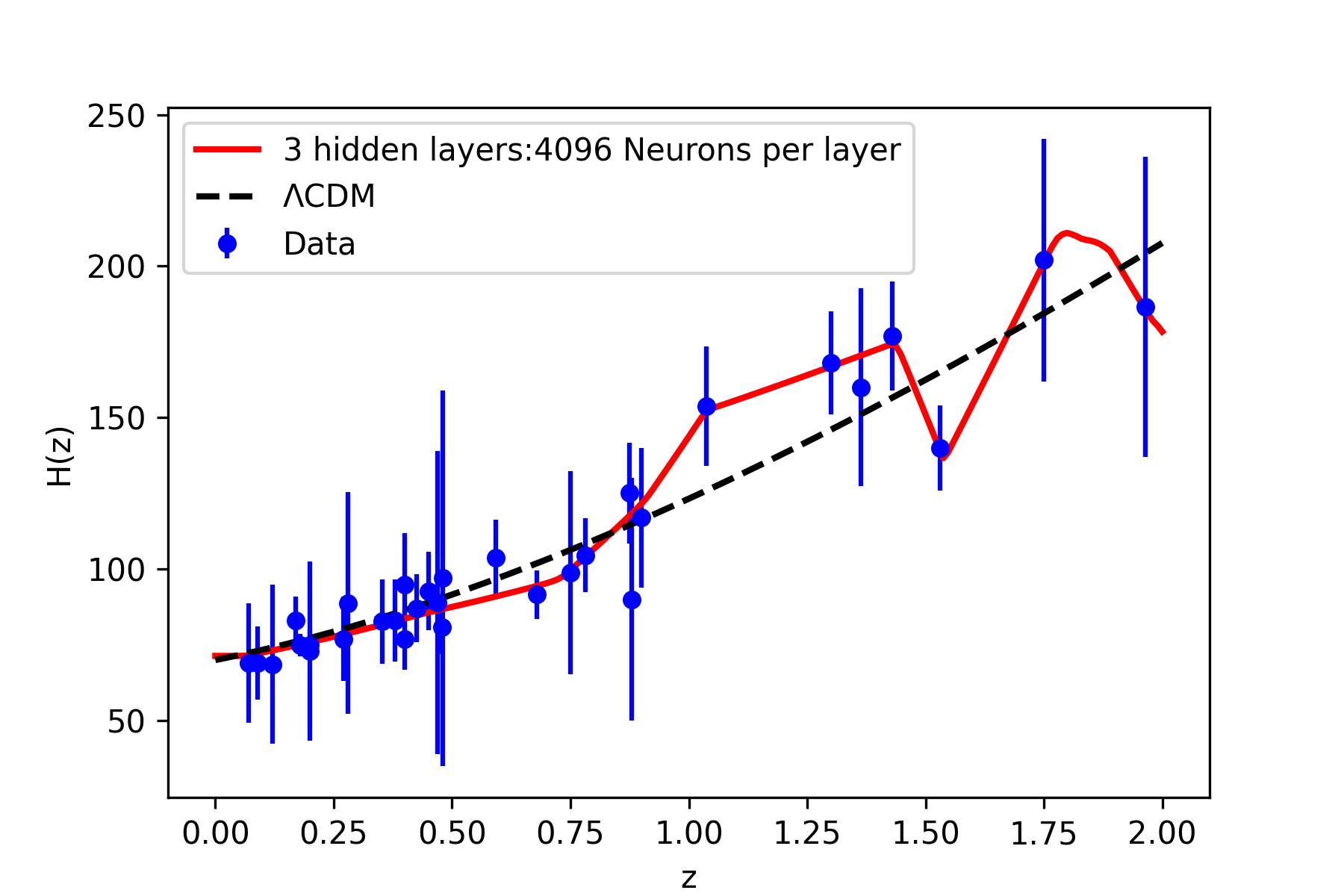}
        (c)
    \end{minipage}
    \hfill
    \begin{minipage}[b]{0.48\textwidth}
        \centering
        \includegraphics[width=\textwidth]{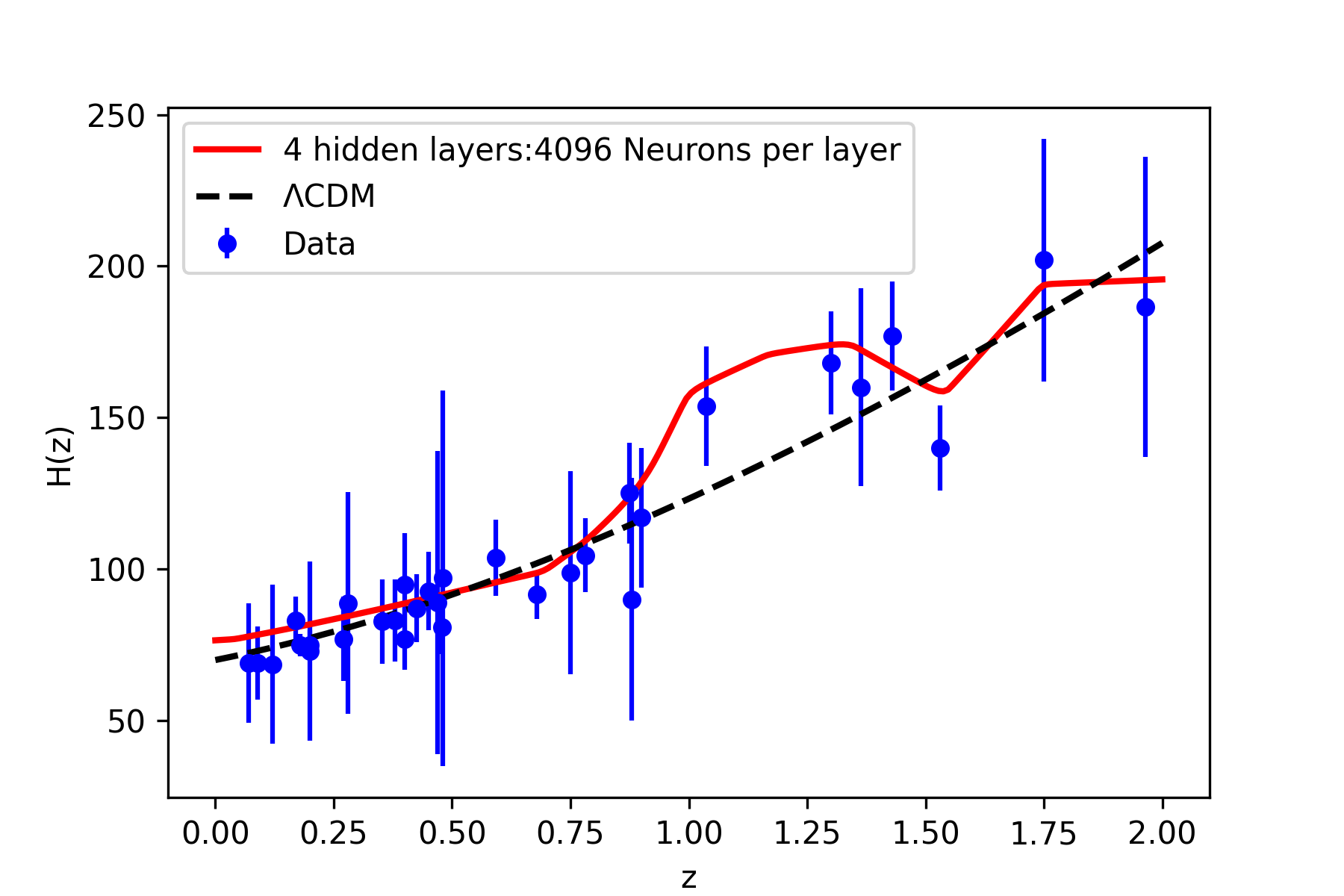}
        (d)
    \end{minipage}
    \caption{Comparison of outputs of  four different ANN models constructed for varying number of hidden layers; with 4096 neurons per layer. We find that ANN with one hidden layer (model a) is sufficient and corresponds to minimum value of the RISK. The outputs of models of b, c, d are overfitted. Adding additional hidden layers does not significantly improve the reconstruction accuracy and instead introduces signs of overfitting.}
\label{fig:layer}
\end{figure*}

Furthermore, the number of layers is varied. Because the Hubble dataset contains only 32 data points, single hidden layer is found to be capable of reconstruction and also corresponded to minimum of RISK. Results for up to four layers are shown in Figure \ref{fig:layer}. So, for the given OHD data, we proceed with a ANN model of one hidden layer with 4096 neurons. \\

We note that the reconstructed $H(z)$ curves become visually stable once the network width exceeds approximately $256$ neurons when bootstrap resampling is applied. This indicates that the ensemble reconstruction already has sufficient representational capacity to capture the underlying expansion history of the observational Hubble data. Increasing the network width beyond this threshold does not significantly alter the reconstructed $H(z)$ curve, although the RISK statistic continues to decrease slightly and reaches its global minimum at $4096$ neurons.\\

\begin{figure}[t]
    \centering
    \includegraphics[width=0.45\textwidth]{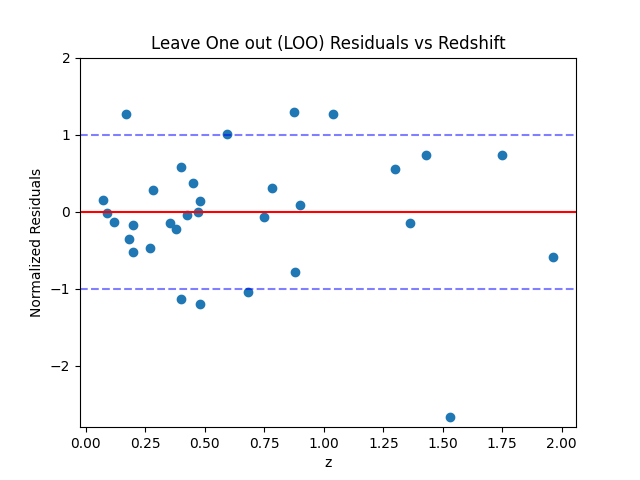}
    \caption{The normalized residuals are distributed around zero with no evident trend as a function of redshift, indicating the absence of systematic bias in the reconstruction or overfitting. The majority of residuals lie within $\pm 1 \sigma$ range, consistent with the expected statistical dispersion of the observational uncertainties. This demonstrates that the model provides a stable and unbiased reconstruction and is not driven by individual data points.}
    \label{fig:loo_resi}
\end{figure}

To assess overfitting we analyse the reconstructions accross different architectures. Networks with additional hidden layers exhibit clear signs of overfitting, while the single layer model produces smooth and stable reconstructions without spurious oscillations. Furthermore, the variance across bootstrap samples remains well behaved, indicating that the model is not fitting noise in the data. For further validation, we perform Leave One Out Cross Validation \citep{Arlot2010, vehtari2017practical}. The results are discussed in Figure \ref{fig:loo_resi}.\\

Smaller networks, however, show evidence of underfitting, which is reflected in higher RISK values and reduced reconstruction accuracy. These results suggest that the chosen architecture lies in a regime that balances flexibility and generalisation.\\

 We, therefore, adopt this configuration, which lies within the stable regime of the reconstruction while maintaining consistency with previous ANN-based cosmological reconstructions.The \textbf{Exponential Linear Unit} (ELU) Function  \cite{Clevert2016} is chosen as the activation function and optimization is performed using the \textbf{Adam} optimizer. The learning rate is chosen to be 0.01 and to be reduced with the number of iterations. The number of optimum epochs is evaluated to be 5750. The optimal ANN hyperparameters for the Hubble parameter reconstruction are summarized in Table \ref{tab:anngrid}.\\

\begin{table}[h!]
\centering
\begin{tabular}{l c}
\hline
\textbf{Parameter} & \textbf{Value} \\
\hline
No. of neurons        & 4096 \\ 
No. of hidden layers  & 1  \\ 
Activation function   & ELU \\ 
Optimizer             & Adam \\  
Loss                  & $\chi^2$\\
\hline
\end{tabular}
\caption{Artificial Neural Network Model for our study}
\label{tab:anngrid}
\end{table}
For training purposes, a loss function is used to tune the weights and biases of the model. Previously, many approaches have been taken including the Mean Absolute Error (MAE) \cite{Wang2019} and the Mean Squared Error (MSE) \cite{GomezVargas2021l}. Later, in their study Huang et al. \cite{Huang2025h} (2025) explored the loss functions and their effect on the calibration further  and combined $\chi^2$ loss with Kullback--Leibler (KL) divergence for calibration. For our study, we have used $\chi^{2}$ as a loss function to be combined with our uncertainty estimation as given by Equation \ref{eq:chi2}.

\begin{equation}
\chi^2_{H} \;=\; \sum_{i=1}^{N} 
\frac{\left[ H_{\mathrm{obs}}(z_i) - H_{\mathrm{pred}}(z_i;\,\theta) \right]^2}
{\sigma_{H,i}^{\,2}} \, .
\label{eq:chi2}
\end{equation}

Given that the OHD dataset contains only 32 measurements, splitting the data into separate training and testing subsets would significantly reduce the effective sample size and limit the information available for model training \cite{Wang2019,Huang2025h}. Instead, we utilise the full dataset during training and employ bootstrap resampling to assess the stability of the model and mitigate overfitting. This approach allows us to exploit the complete information content of the dataset while still providing a robust estimate of model performance.\\

\begin{figure}[h]
    \centering
    \includegraphics[width=0.45\textwidth]{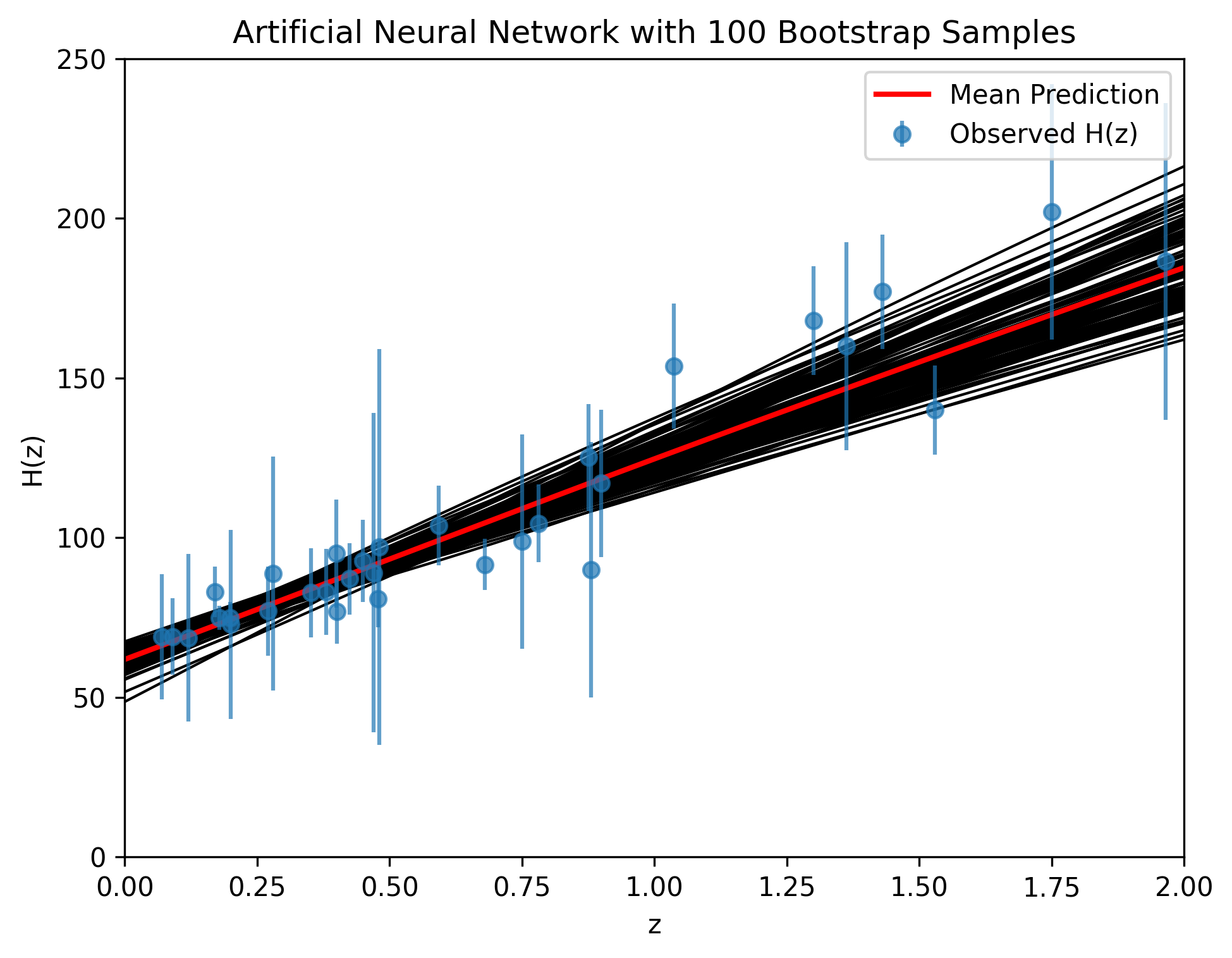}
    \caption{Artificial Neural Network predictions obtained from running the network for 100 different bootstrap samples from the OHD dataset.
Each curve corresponds to a reconstruction derived from an ANN. The variation in prediction is expected from a stochastic process.}
    \label{fig:bt100}
\end{figure}

There's two different sources of uncertainty  to be taken into account in this context. Observational uncertainties in the OHD data correspond to aleatoric uncertainty, arising from measurement errors, while uncertainties related to the model reconstruction reflect epistemic uncertainty, associated with limited data and model flexibility.\\

ANNs do not inherently provide uncertainty estimates related to the reconstruction. There have been approaches in previous studies to mitigate the problem. Chen et al.(2025) \cite{Chen2025} trained  two separate neural networks to model $H(z)$ and its associated uncertainty, whereas Shah et al. (2024) \cite{Shah2024} employed the Kullback-Leibler divergence as a loss function to preserve the physical interpretation of $H(z)$ and its uncertainty. In a complementary approach, Mukherjee et al.  \citep{mukherjee2026} propagate uncertainties by combining ANN-reconstructed luminosity distance errors, including covariance and an intrinsic scatter term within a Gaussian MCMC likelihood framework.\\

To account for sampling variability and assess the robustness of the reconstruction, we supplement the ANN model with a bootstrap procedure. Bootstrap resampling draws datasets of the same size as the original OHD sample by sampling with replacement, allowing individual data points to appear multiple times within a resample \citep{efron1993bootstrap}. Each bootstrap realization is used to train the network independently, producing a distribution of reconstructed $H(z)$ functions. The bootstrap realisations here can be considered as ensembles and the final reconstruction as the ensemble average. We generate 1000 bootstrap samples, and the final reconstruction is obtained by averaging over these realizations, with the variance providing an estimate of the reconstruction uncertainty.\\

The bootstrap resampling procedure employed in the ANN framework primarily captures the variability induced by sampling fluctuations in the dataset and, therefore, provides a representation of aleatoric uncertainty. As a result, ANN-based uncertainty estimates should be interpreted as an approximate measure of reconstruction variance rather than a complete characterization of model uncertainty.\\

\begin{figure}[b]
    \centering
    \includegraphics[width=0.45\textwidth]{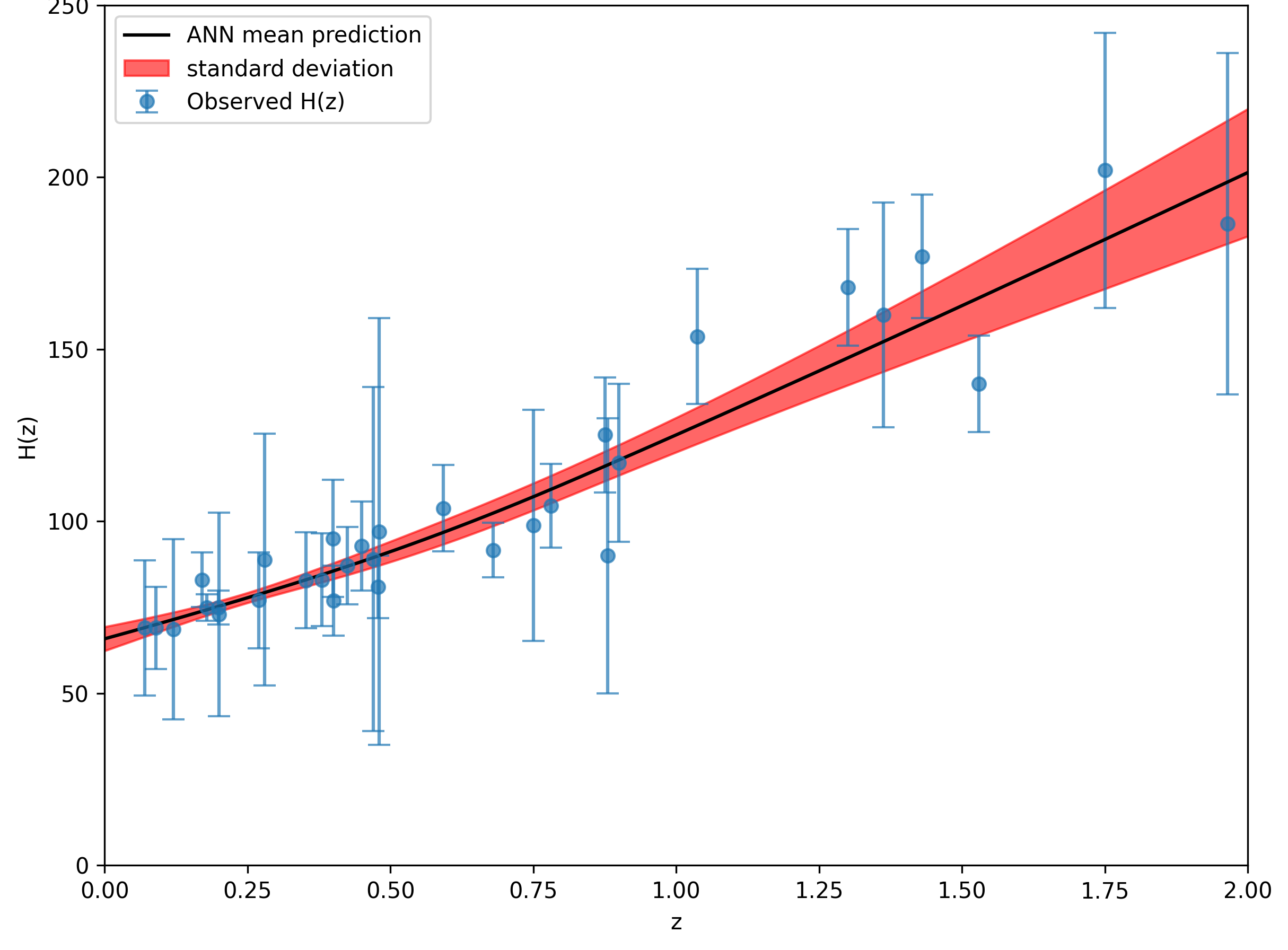}
    \caption{Reconstruction of the Hubble parameter $H(z)$ using an Artificial Neural Network (ANN) framework. 
The black points with error bars represent the OHD, while the solid curve corresponds to the mean ANN prediction. 
The shaded region indicates the $1\sigma$ uncertainty interval derived from bootstrap realizations of the training dataset.}
    \label{fig:annf}
\end{figure}

Figure \ref{fig:bt100} shows an example of the bootstrap prediction distribution with 100 samples chosen randomly from our study. The final predictions with uncertainty intervals are illustrated in Figure \ref{fig:annf}. For our analysis, we utilised the PyTorch library \cite{Paszke2019}.
\subsection{Bayesian Neural Network Reconstruction of H(z)}

\label{subsec:bnn}

We adopt a Bayesian Neural Network (BNN) framework to explicitly model uncertainties in the reconstruction of the Hubble parameter. As compared to conventional neural networks, which train by point estimation, BNNs treat the network weights as random variables and infer their posterior distributions conditioned on the observed data \citep{Neal1996,MacKay1995}.\\

This enables propagation of uncertainty from the model parameters to the predicted observables capturing both aleatoric and epistemic uncertainty arising from limited or noisy datasets \citep{Gelman2013,ghahramani2015probabilistic}. This distinction highlights the advantage of the Bayesian framework in providing a more complete characterization of uncertainty compared to the ANN-based approach. This formulation of neural networks were originally developed by Bishop et al. \citep{bishop2006prml,bishop2013brml}, and have since become a standard tool for uncertainty inference. Given the observed data $\mathcal{D}$, the posterior distribution of the parameters is obtained via Bayes' theorem,
\begin{equation}
p(\mathbf{w} \mid \mathcal{D}) = \frac{p(\mathcal{D} \mid \mathbf{w})\, p(\mathbf{w})}{p(\mathcal{D})},
\end{equation}
where $p(\mathcal{D} \mid \mathbf{w})$ denotes the likelihood and $p(\mathcal{D})$ is the probability of observing the data. $p(\mathrm{w})$ denotes the prior of the weights. BNN reduces the training to a probabilistic parameter inference problem.\\

Assuming Gaussian observational uncertainties, the likelihood function is written as
\begin{equation}
p(\mathbf{y} \mid \mathbf{x}, \mathbf{w}) \propto 
\exp\left[
-\frac{1}{2}\sum_i
\frac{\left(y_i - f(\mathbf{x}_i;\mathbf{w})\right)^2}{\sigma_i^2}
\right],
\end{equation}
where $f(\mathbf{x};\mathbf{w})$ denotes the network output. The assumption of Gaussian observational uncertainties is motivated by the central limit theorem and the form of reported measurement errors in the OHD  dataset \citep{Ratra2023}.\\

However, recent studies have shown that this assumption may not always hold, particularly in heterogeneous or high-redshift datasets, where deviations from Gaussianity can arise due to systematic effects and intrinsic scatter \citep{dainotti2024,Favale:2024lgp}. In such cases, the Gaussian likelihood may not fully capture the tails of the error distribution, potentially leading to under or overestimation of uncertainties in the reconstructed quantities. The BNN framework partially mitigates this limitation by marginalizing over the posterior distribution of the network parameters, thereby providing a more flexible characterization of uncertainty compared to deterministic approaches.\\
 
The dimensionality of the Bayesian inference is given by the total number of network parameters included in the posterior. Due to the high dimensionality and nonlinearity Bayesian inference over neural network parameters is  not analytically tractable \citep{Neal1996}.\\

For inference of the posterior distribution, two broad computational approaches are commonly employed: sampling-based methods such as Markov Chain Monte Carlo (MCMC)\cite{neal2011mcmc,Gelman2013}, and approximate methods such as variational inference \citep{blei2017variational}. Although variational approaches are computationally efficient, sampling-based techniques provide a more faithful characterization of the posterior distribution \citep{ghahramani2015probabilistic,blundell2015weight}. Model predictions are obtained by marginalising over the posterior,
\begin{equation}
p(y^{*} \mid x^{*}, \mathcal{D}) =
\int p(y^{*} \mid x^{*}, \mathbf{w})\, p(\mathbf{w} \mid \mathcal{D})\, \mathrm{d}\mathbf{w},
\end{equation}
which yields both predictive means and associated uncertainties.\\

In the Bayesian framework, prior distributions are assigned to all network parameters (weights and biases), which are then subsequently updated through the likelihood informed by the data. Following established treatments \citep{MacKay1991,MacKay1992b,Neal1996}, we impose independent zero-mean Gaussian priors , $
\mathcal{N}(0, \sigma^2)
$, on the weights and biases of the network, thus treating positive and negative weights as equally probable \citep{MacKay1992,Hinton1993}. This reflects the absence of any preferred parameter values prior to observing the data.\\

\begin{figure}[b]
    \centering
    \includegraphics[width=0.45\textwidth]{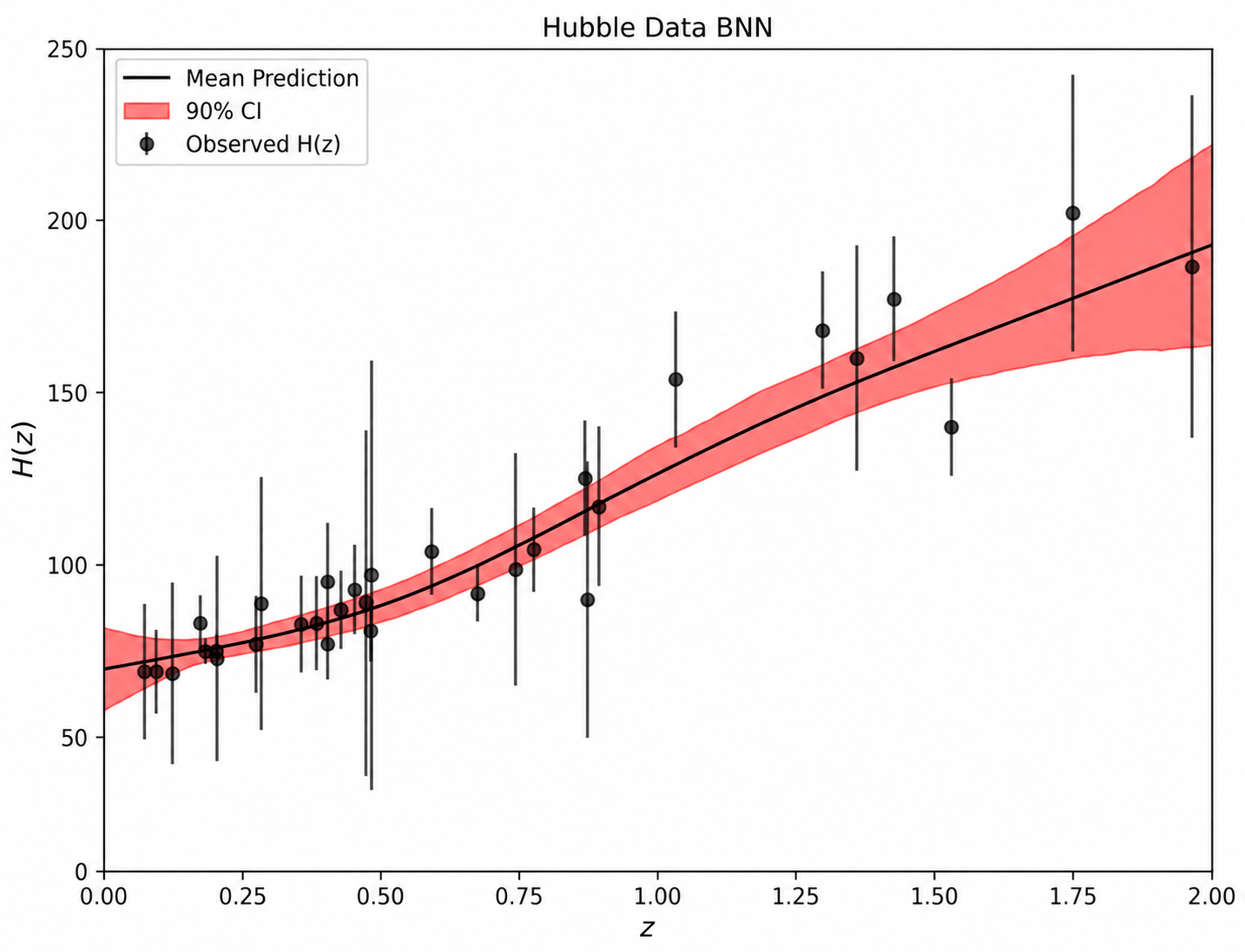}
    \caption{
Reconstruction of the Hubble parameter $H(z)$ using a Bayesian Neural Network (BNN). 
The black points with error bars denote the OHD, while the solid curve represents the mean prediction of the BNN model. 
The shaded region corresponds to the $1\sigma$ credible interval obtained from the posterior distribution of the network parameters, illustrating the uncertainty propagation inherent to the Bayesian framework.
} 
    \label{fig:bnnf}
\end{figure}
Model comparison across the explored hyperparameter configurations is performed using the Widely Applicable Information Criterion (WAIC), which estimates out-of-sample predictive accuracy while penalizing model complexity \cite{watanabe2010asymptotic,vehtari2017practical}. The WAIC is defined as:
\begin{equation}
\mathrm{WAIC} = -2\,\mathrm{ELPD}_{\mathrm{WAIC}}.
\end{equation}

\noindent Where expected log predictive density $ELPD_{WAIC}$ is :
\begin{equation}
\mathrm{ELPD}_{\mathrm{WAIC}} = \mathrm{lppd} - p_{\mathrm{WAIC}},
\end{equation}

\begin{equation}
\mathrm{lppd} = \sum_{i=1}^{N} 
\log \left( \frac{1}{S} \sum_{s=1}^{S} p(y_i \mid \theta^{(s)}) \right),
\end{equation}

and the effective number of parameters is

\begin{equation}
p_{\mathrm{WAIC}} =
\sum_{i=1}^{N} 
\mathrm{Var}_{s}\left(\log p(y_i \mid \theta^{(s)})\right).
\end{equation}

In this framework, ($\mathrm{ELPD}_{\mathrm{WAIC}}$) quantifies the predictive performance of the model, while the effective number of parameters ($p_{\mathrm{WAIC}}$) measures the inferred functional complexity. WAIC is computed for each combination of network width and prior variance to evaluate the balance between predictive accuracy and model flexibility. \\
\begin{table}[h]
\centering
\begin{tabular}{ccc}
\hline
Neurons & Standard Deviation $\sigma$ & WAIC \\
\hline
64 & 5.0 & 260.912 \\
16 & 5.0 & 261.884 \\
32 & 5.0 & 261.984 \\
32 & 10.0 & 262.732 \\
64 & 2.0 & 262.74 \\
\hline
\end{tabular}
\caption{Best performing BNN models ranked by WAIC for different hidden neuron sizes and prior standard deviations. }
\label{tab:waic_models}
\end{table}

We find that the predictive performance is largely insensitive to the network width once sufficient capacity is reached, whereas the prior variance plays the dominant role in regulating the effective model complexity. The best-performing models based on the ELPD-WAIC criterion are summarized in Table~\ref{tab:waic_models}. The determined parameters for the BNN model are summarized in Table~\ref{tab:bnnmodel}.\\

\begin{table}[h!]
\centering
\begin{tabular}{lc}
\hline
Parameter & Value \\
\hline
Hidden layers & 1 \\
Neurons per layer & 64 \\
Prior &(0,$5^2$)\\
Activation function & ELU \\
\hline
\end{tabular}
\caption{Optimum Parameters for the Bayesian Neural Network Model}
\label{tab:bnnmodel}
\end{table}

In this work, we adopt MCMC methods to train our BNNs, motivated by the relatively shallow network architecture which render MCMC computationally feasible. An important advantage of sampling-based inference is that, in the limit of a large number of samples, the generated chain asymptotically converges to the true posterior distribution.
We employed the No-U-Turn Sampler (NUTS) as introduced by Hoffman et al. \cite{hoffman2011}, an adaptive variant of Hamiltonian Monte Carlo (HMC), as supplement. HMC explores the parameter space by first introducing auxiliary momentum variables. Then evolving the system according to Hamiltonian dynamics which enables efficient sampling even in case of high-dimensional spaces.Traditional HMC requires manual tuning of the integration step size and trajectory length. The NUTS algorithm eliminates this requirement by adaptively selecting these parameters during the warm-up phase. This automates the sampling procedure and improves sampling efficiency.\\

\begin{figure*}
    \centering
    \includegraphics[width=0.7\textwidth]{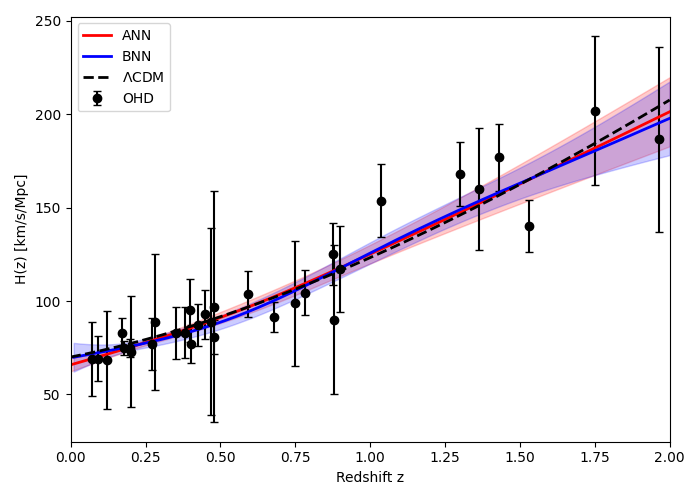}
    \caption{
Reconstruction of the Hubble parameter $H(z)$ obtained using Artificial Neural Network (ANN) and Bayesian Neural Network (BNN) frameworks. 
The black points with error bars represent the OHD, while the dashed curve corresponds to the $\mathrm{\Lambda}$CDM prediction with $H_0=70\,\mathrm{km\,s^{-1}\,Mpc^{-1}}$ and $\Omega_m=0.3$. 
The close agreement between the ANN and BNN reconstructions demonstrates the robustness of the neural-network calibration.
}
    \label{fig:combo}
\end{figure*}

The resultant reconstruction of Hubble parameter with uncertainties from the posterior is illustrated in Figure \ref{fig:bnnf}. Figure \ref{fig:combo} shows the $H(z)$ reconstructions obtained from ANN and BNN compared with standard $\mathrm{\Lambda}$CDM.  The Pyro interface by Bingham et al. \cite{Bingham2019} is used for the purpose.  \\

\subsection{Constraints on the Amati Relation Parameters}
\label{subsec:amati}
Following the reconstruction of the Hubble Data using the NN frameworks, the luminosity distance of the GRBs are calculated. Using $H(z)$, the luminosity distance $d_L(z)$ is obtained by integrating the inverse expansion rate : 
\begin{equation}
d_L(z) = (1+z)\, c \int_0^z \frac{\mathrm{d}z'}{H(z')},
\end{equation}
where $c$ is the speed of light.\\

The isotropic-equivalent radiated energy $E_{\mathrm{iso}}$ of a gamma-ray burst is computed using the bolometric fluence $S_{\mathrm{bolo}}$ as
\begin{equation}
E_{\mathrm{iso}} = 4 \pi d_L^2(z)\, \frac{S_{\mathrm{bolo}}}{1+z}.
\end{equation}
The bolometric fluence is expressed in units of $\mathrm{erg\,cm^{-2}}$.
The observed spectral peak energy $E_{\mathrm{peak}}^{\mathrm{obs}}$ is shifted by cosmic expansion.
The corresponding rest-frame peak energy is calculated using redshift correction term -
\begin{equation}
E_{\mathrm{peak}} = E_{\mathrm{peak}}^{\mathrm{obs}} (1+z).
\end{equation}\\

The Amati relation \citep{Amati2002, Amati2008} links the isotropic-equivalent energy to the rest-frame spectral peak energy via a power-law relation of the form

\begin{equation}
y_i = a + b\,x_i
\end{equation}
where
\begin{equation}
x_i = \log\!\left(\frac{E_{p,i}}{300~\text{keV}}\right), 
\qquad
y_i = \log\!\left(\frac{4\pi d_L^2 S_{\mathrm{bolo},i}}{1+z}\right).
\end{equation}
Here, $a$ and $b$ are Amati calibration parameters determined from the data.\\

For the propagation of errors related to $x_i$, the error in terms of peak energy error term is calculated as 
\begin{equation}
\sigma_{x_i} = \frac{\sigma_{E_p}}{\ln(10)\, E_{p,\mathrm{rest}}}
\end{equation}

\noindent For the errors on the $y_i$ term, first the error on the isotropic energy is calculated using error propagation -
 \begin{equation}
\sigma_{E_{\mathrm{iso}}}
= E_{\mathrm{iso}}
\sqrt{
\left( \frac{2\,\sigma_{d_L}}{d_L} \right)^2
+
\left( \frac{\sigma_{S_{bolo}}}{S_{\mathrm{bolo}}} \right)^2
}
\end{equation}
Then the error on $y_i$ term is obtained equivalently as
\begin{equation}
\sigma_{y_i}
= \frac{\sigma_{E_{\mathrm{iso}}}}{\ln(10)\, E_{\mathrm{iso}}}
\end{equation}\\

The likelihood function for the standard Amati relation is  then defined as a function of the Amati Parameters a, b and $\sigma_{tot}$ as
\begin{equation}
\mathcal{L} \propto 
\prod_{i=1}^{N}
\frac{1}{\sqrt{2\pi\,\sigma_{\mathrm{tot},i}^2}}
\exp\left[
-\frac{\left(y_i - a - b x_i\right)^2}
{2\,\sigma_{\mathrm{tot},i}^2}
\right],
\end{equation}

\noindent The total variance is given by  -
\begin{equation}
\sigma_{\mathrm{tot}}^2 = \sigma_{\mathrm{ext}}^2 + \sigma_y^2 + b^2\sigma_x^2.
\end{equation}
here $\sigma_{ext}$ refers to the intrinsic scatter of the energy relation which is independent of measurement errors and is constrained as a model parameter.\\

\begin{figure*}
\centering

\begin{minipage}{0.48\textwidth}
\centering
\includegraphics[width=\textwidth]{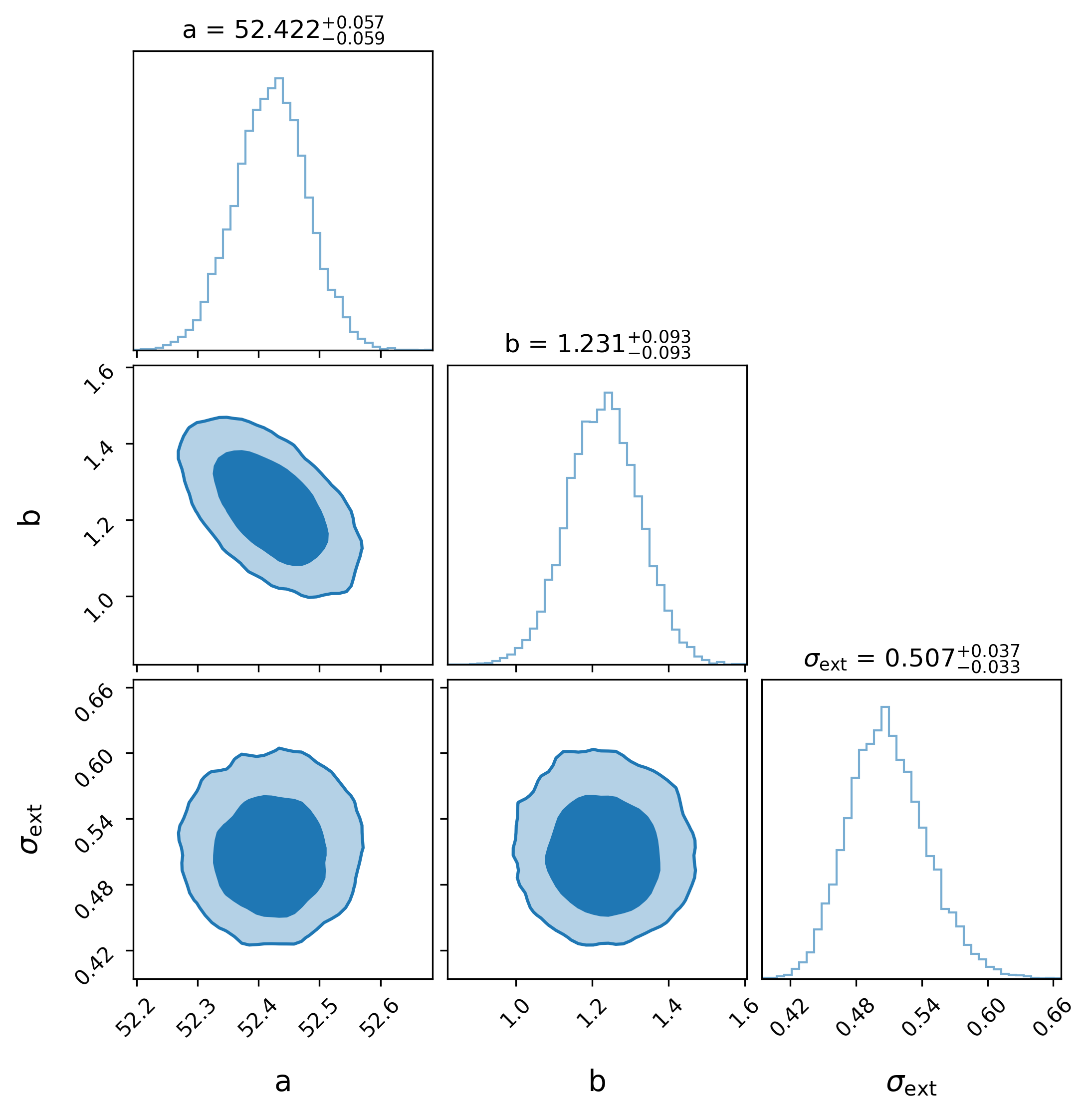}
(a) ANN
\end{minipage}
\hfill
\begin{minipage}{0.48\textwidth}
\centering
\includegraphics[width=\textwidth]{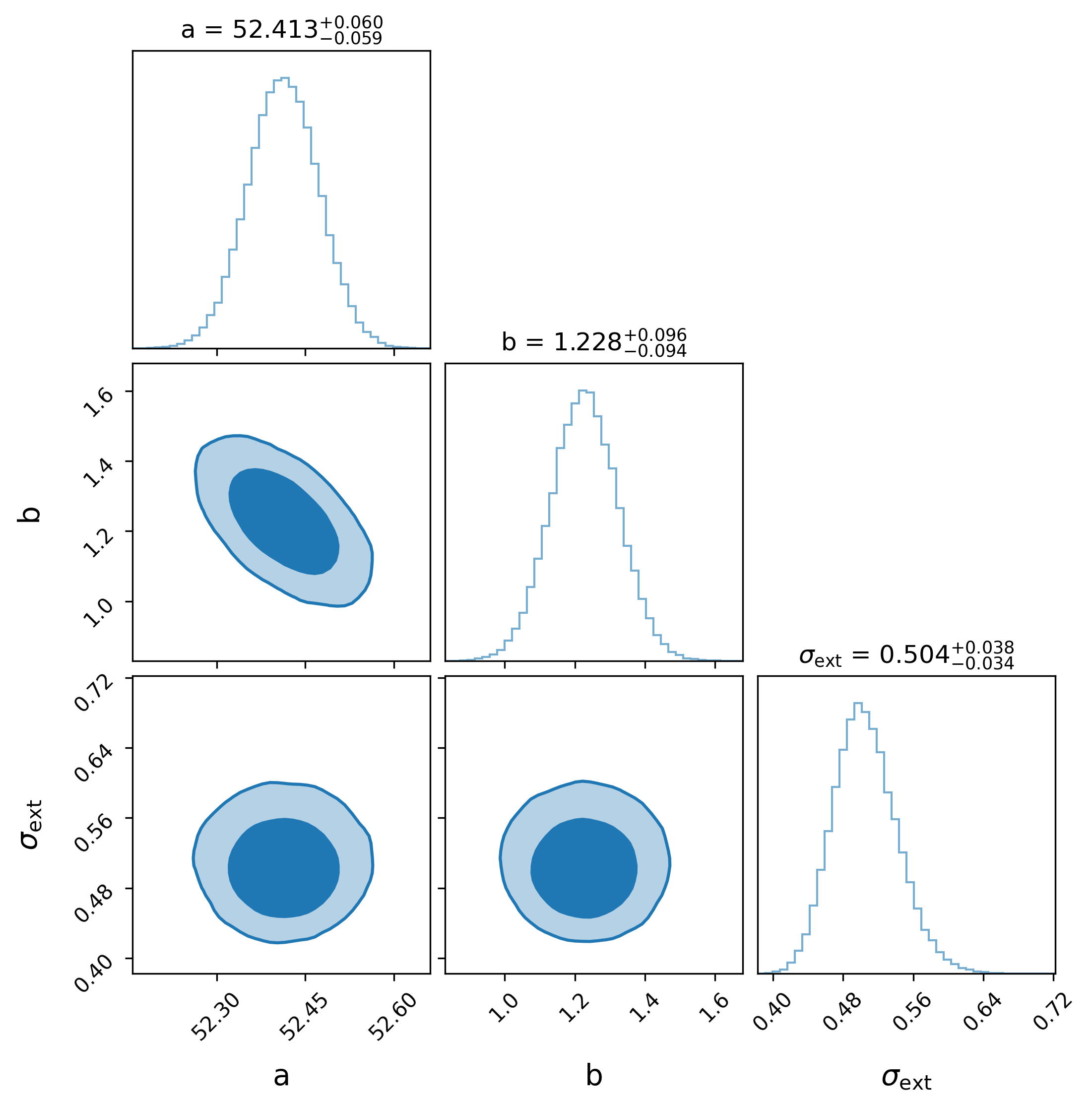}
(b) BNN
\end{minipage}

\caption{Corner plots of the posterior distributions for the Amati parameters
$(a, b, \sigma_{\rm ext})$ obtained from the A220 dataset using ANN and BNN models.
The contours indicate the 68\% and 95\% credible regions, while the diagonal panels
show the marginalised one-dimensional posterior distributions of the parameters.}

\label{fig:a220_methods}

\end{figure*}

\begin{figure*}
\centering

\begin{minipage}{0.48\textwidth}
\centering
\includegraphics[width=\textwidth]{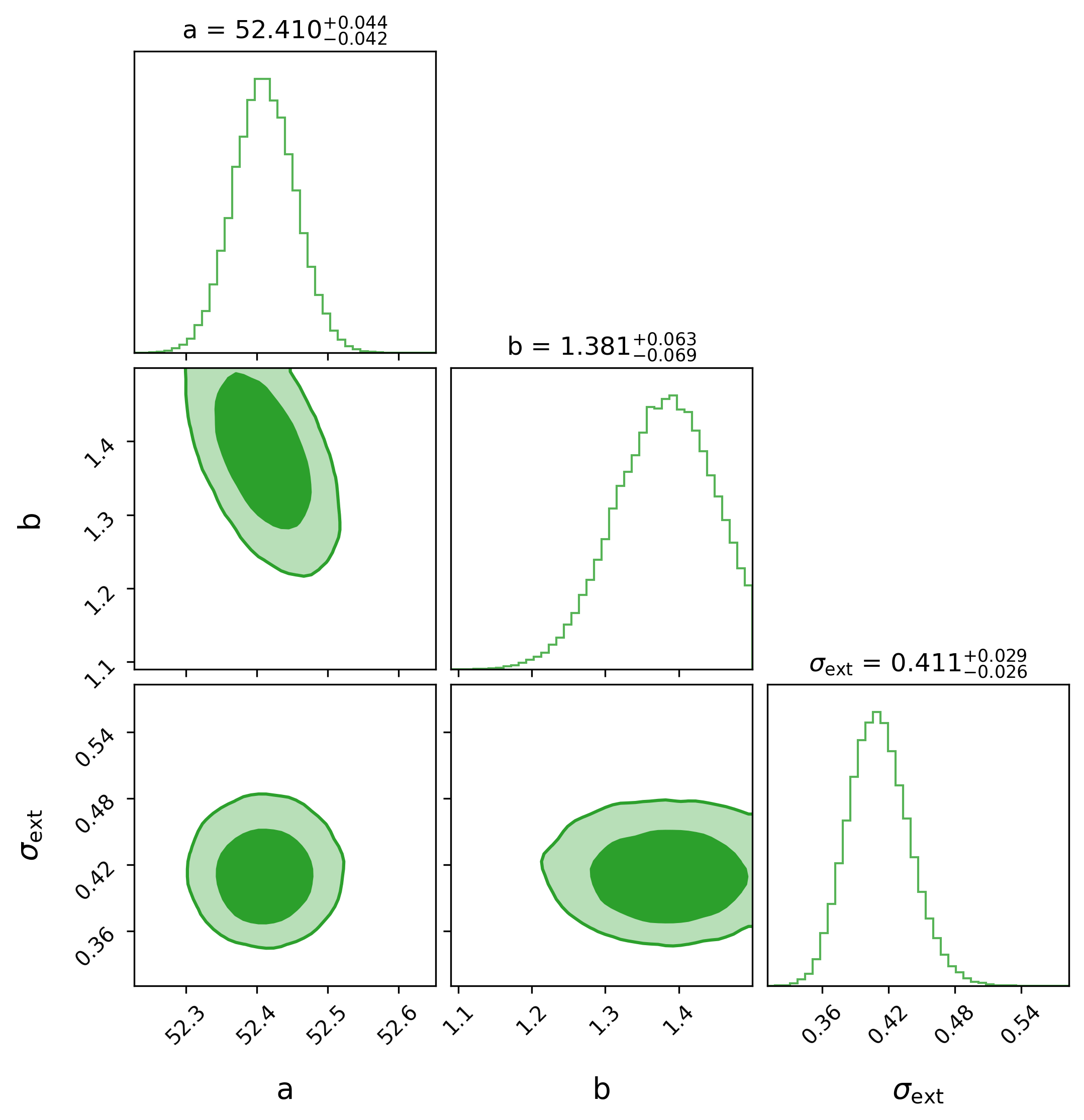}
(a)ANN
\end{minipage}
\hfill
\begin{minipage}{0.48\textwidth}
\centering
\includegraphics[width=\textwidth]{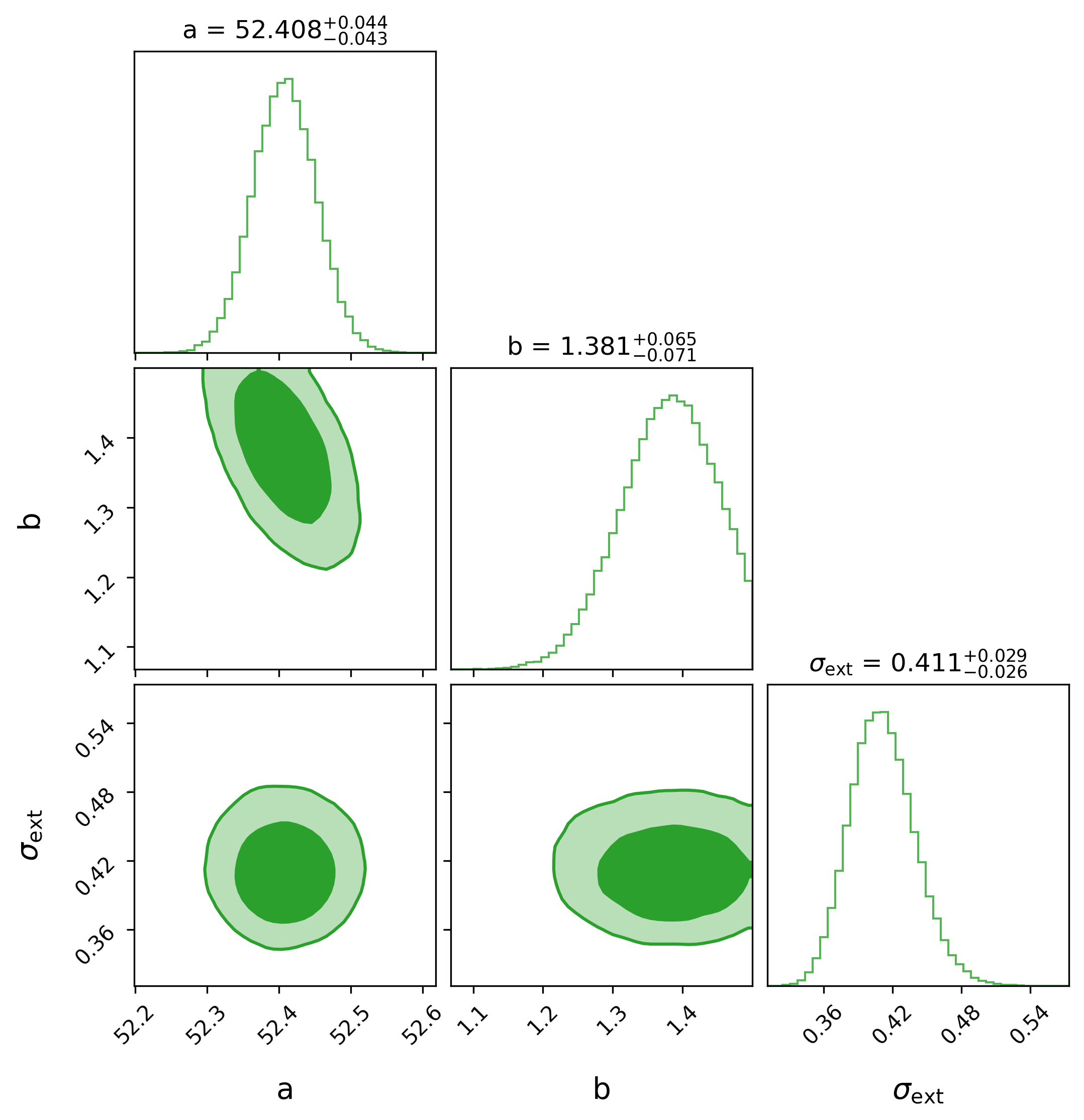}
(b) BNN
\end{minipage}

\caption{Corner plots of the posterior distributions for the Amati parameters
$(a, b, \sigma_{\rm ext})$ obtained from the J220 dataset using ANN and BNN models.
The contours indicate the 68\% and 95\% credible regions, while the diagonal panels
show the marginalised one-dimensional posterior distributions of the parameters.
The reduced intrinsic scatter of the J220 sample suggests a more homogeneous GRB
population or reduced observational systematics compared with the A220 dataset.}

\label{fig:j220_methods}

\end{figure*}

We perform Markov Chain Monte Carlo (MCMC) sampling using the \texttt{emcee} affine-invariant ensemble sampler \citep{ForemanMackey2013} to simultaneously constrain the parameters (a), (b), and the intrinsic scatter ($\sigma_{\mathrm{ext}}$) for the ANN and BNN reconstructed data. The parameter space is explored using 32 walkers evolved for 5000 steps, with the first 1000 steps discarded as burn-in and the chains thinned by a factor of 10 to reduce autocorrelation. Uniform priors are adopted in the ranges (50 < a < 60), (0 < b < 2), and (0 < $\sigma_{\mathrm{ext}}$ < 5). The posterior distributions are obtained from the flattened chains, and the best-fit parameter values are taken as the median of the posterior samples. The resulting posterior distributions and best-fit values are shown in Figures \ref{fig:a220_methods} and \ref{fig:j220_methods}, respectively.\\

\begin{figure*}
\centering

\begin{minipage}{0.48\textwidth}
\centering
\includegraphics[width=\textwidth]{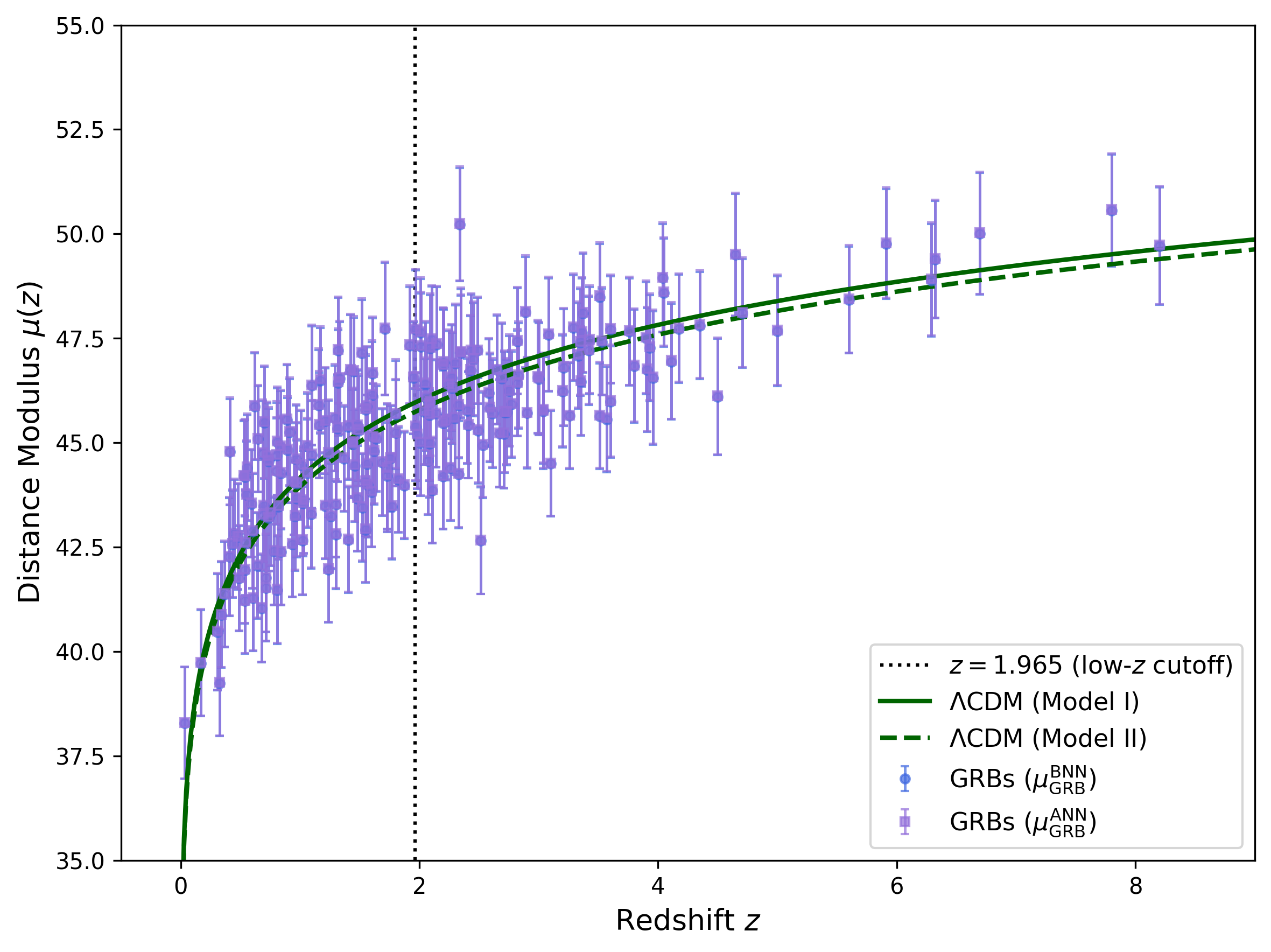}
(a)A220
\end{minipage}
\hfill
\begin{minipage}{0.48\textwidth}
\centering
\includegraphics[width=\textwidth]{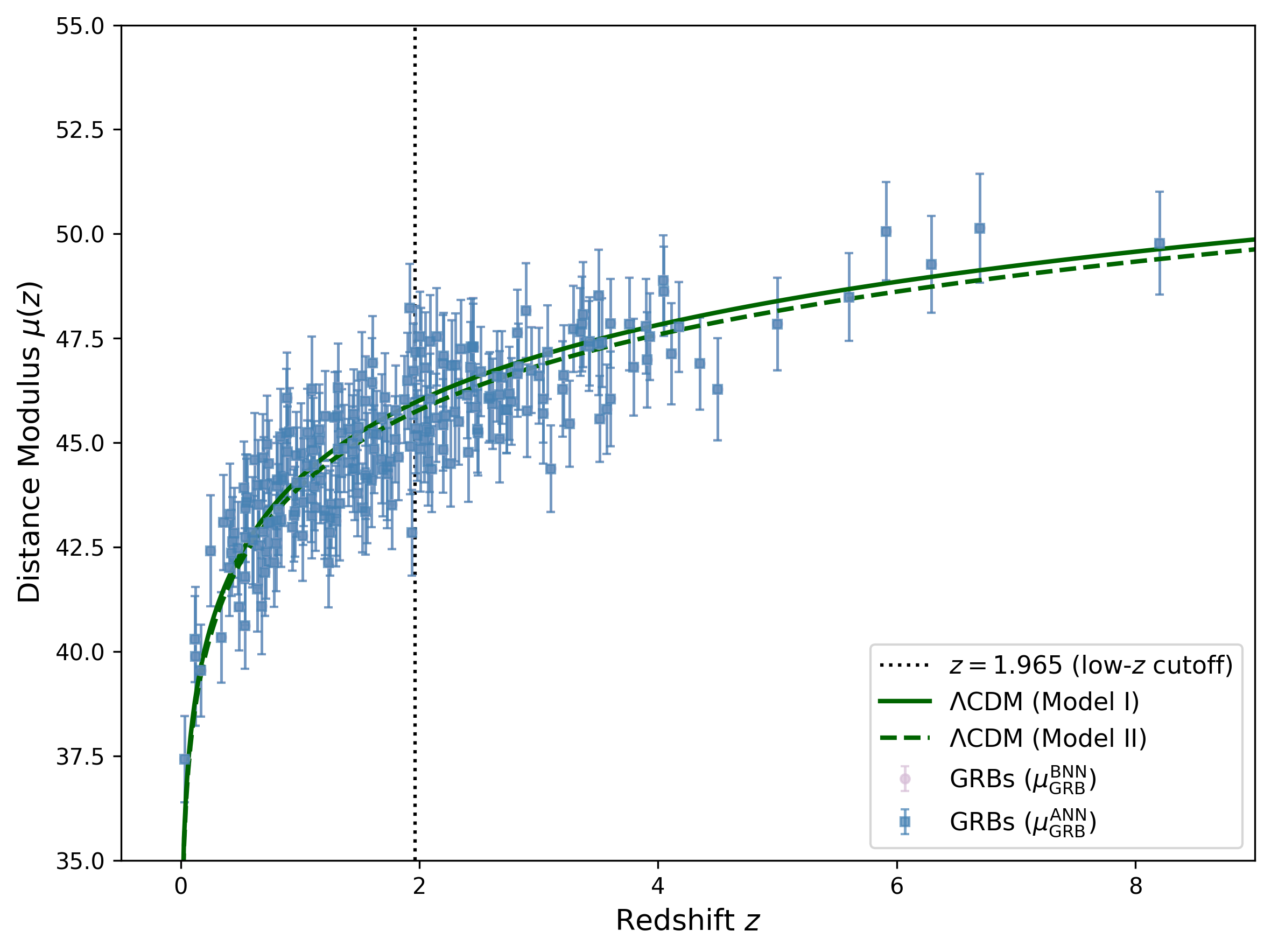}
(b) J220
\end{minipage}

\caption{Hubble diagram of GRBs for the A220 (left) and J220 (right) samples. GRBs with $z \leq 1.965$ are calibrated using OHD data via ANN reconstruction, while those with $z > 1.965$ are derived from the Amati relation. The blue and purple points show the GRB distance moduli reconstructed using BNN and ANN methods, respectively. The vertical dashed line marks the low-$z$ cutoff at $z = 1.965$. The solid and dashed curves represent the $\mathrm{\Lambda}$CDM distance modulus corresponding to cosmological parameters from Planck CMB measurements $H_0 = 67.36,\mathrm{km,s^{-1},Mpc^{-1}},\ \Omega_m = 0.315$ and Pantheon+ supernova constraints $H_0 = 73.6,\mathrm{km,s^{-1},Mpc^{-1}},\ \Omega_m = 0.334$, respectively.
}

\label{fig:hubble_grb}

\end{figure*}

The results obtained from the calibration are used to estimate the luminosity distances of GRBs at higher redshifts.
The corresponding distance modulus for GRBs is obtained as

\begin{equation}
\begin{split}
\mu_{\rm GRB} =
\frac{5}{2}
\left[
\log \left(\frac{E_{\rm iso}}{1\,{\rm erg}}\right)
-\log \left(\frac{4\pi}{1+z}\right)
\right. \\
\left.
-\log \left(\frac{S_{\rm bolo}}{1\,{\rm erg\,cm^{-2}}}\right)
\right]
-97.45 ,
\end{split}
\end{equation}

\noindent which are then used to construct the GRB Hubble diagram as shown in Figure~\ref{fig:hubble_grb}. 

\section{Results}
\label{sec:results}
\begin{table}[h]
\centering
\textbf{A220}\smallskip
\label{tab:a220_comparison}

\begin{tabular}{lccc}

\hline\noalign{\smallskip}
\textbf{Method} & \textbf{$a$} & \textbf{$b$} & \textbf{$\sigma_{\rm ext}$} \\
\noalign{\smallskip}\hline\noalign{\smallskip}

$ANN^a$
& $52.422^{+0.057}_{-0.059}$
& $1.231^{+0.093}_{-0.093}$
& $0.507^{+0.037}_{-0.033}$ \\

$BNN^a$
& $52.413^{+0.060}_{-0.059}$
& $1.228^{+0.096}_{-0.094}$
& $0.504^{+0.038}_{-0.034}$ \\

Liang et al.\cite{Liang2022}
& 
& $1.298^{+0.090}_{-0.080}$
& $0.511^{+0.047}_{-0.047}$ \\

Liu et al. \cite{Liu:2022inf}
&
& $1.290^{+0.126}_{-0.126}$
& $0.521^{+0.037}_{-0.034}$ \\

Huang et al.\cite{Huang2025h}
& 
& $1.28^{+0.13}_{-0.13}$
& $0.52^{+0.03}_{-0.04}$ \\

\noalign{\smallskip}\hline
\end{tabular}

\caption{Comparison of Amati relation parameters ($a$, $b$, $\sigma_{\rm ext}$) for the A220 dataset. 
$^{a}$Results obtained in this work using neural-network calibration methods. While we tabulate the values of b and $\sigma_{ext} $,  the values of a aren't tabulated due to different choice of formulation for Amati Relation across the papers.}
\label{tab:a220pramati}
\end{table}

We find that for the J220 sample at $z < 1.965$, the neural-network-based calibration yields a higher slope and reduced intrinsic scatter compared to the corresponding A220 subsample. The smaller intrinsic scatter suggests a tighter correlation within the J220 dataset, indicating a more homogeneous GRB population or reduced observational systematics. The steeper slope further implies a stronger dependence between the observables within this subsample. Such differences between subsamples may arise from variations in sample selection, redshift coverage, or measurement uncertainties, which can affect the apparent strength and dispersion of the empirical relation. The consistency of the inferred parameters across different inference methods supports the robustness of this behaviour.\\

For the A220 sample, we obtain a slope of 
$b = 1.231^{+0.093}_{-0.093}$ using the Artificial Neural Network (ANN) approach and 
$b = 1.218^{+0.096}_{-0.094}$ using the Bayesian Neural Network (BNN) framework. 
These estimates are mutually consistent and agree, within 1 $\sigma$ uncertainties, with previous low-redshift calibrations obtained using alternative model-independent techniques, including 
$b = 1.298^{+0.090}_{-0.080}$ and $\sigma_{ext} = 0.511^{+0.047}_{-0.047}$ from Gaussian-process reconstruction \cite{Liang2022} and 
$b = 1.290^{+0.126}_{-0.126}$ and $\sigma_{ext} = 0.521^{+0.037}_{-0.034}$  obtained through interpolation methods \cite{Liu:2022inf}. 
They are also consistent with the ANN-based calibration yielding 
$b = 1.28^{+0.13}_{-0.13}$ and $\sigma_{ext} = 0.52^{+0.03}_{-0.04}$ \cite{Huang2025h}, while our results provide a slightly tighter constraint on the Amati slope.\\
\begin{table}[h!]
\centering

\begin{tabular}{l c c c}
\hline
Model & $a$ & $b$ & $\sigma_{\rm ext}$ \\
\hline

ANN$^{a}$ 
& $52.410^{+0.044}_{-0.042}$ 
& $1.381^{+0.063}_{-0.069}$ 
& $0.411^{+0.029}_{-0.026}$ \\

BNN$^{a}$ 
& $52.408^{+0.044}_{-0.043}$ 
& $1.381^{+0.065}_{-0.071}$ 
& $0.411^{+0.029}_{-0.026}$ \\

Cao et al.~\cite{Cao:2024vmo}
& 
& $1.379^{+0.087}_{-0.087}$
& $0.382$ \\

Jia et al.~\cite{Jia2022}
& 
& $1.46^{+0.06}_{-0.06}$
& $0.39^{+0.02}_{-0.02}$ \\

Huang et al.~\cite{Huang2025h}
& 
& $1.52^{+0.09}_{-0.09}$
& $0.42^{+0.03}_{-0.03}$ \\

\hline
\end{tabular}

\caption{Amati parameter estimations ($a$, $b$, $\sigma_{\rm ext}$) for the J220 dataset. 
$^{a}$Results obtained in the present work using neural-network calibration methods.}

\label{tab:j220pramati}
\end{table}

Table~\ref{tab:j220pramati} summarizes the results for the J220 sample. We obtain a slope of 
$b = 1.381^{+0.063}_{-0.069}$ from the ANN calibration and 
$b = 1.381^{+0.065}_{-0.071}$ from the BNN calibration. 
The close agreement between the ANN- and BNN-based estimates indicates that the inferred correlation parameters are robust with respect to the choice of neural network framework. 
These values are broadly consistent with results obtained through simultaneous fitting of cosmological and correlation parameters within a $\mathrm{\Lambda}$CDM framework, 
$b = 1.379^{+0.087}_{-0.087}$ and $\sigma_{ext} = 0.382$ \cite{Cao:2024vmo}, as well as with results obtained by Redshift-binning method \cite{Jia2022}, 
$b = 1.46^{+0.06}_{-0.06}$ and  $\sigma_{ext} = 0.39^{+0.02}_{-0.02}$.  
Compared to the ANN-based reconstruction yielding 
$b = 1.52^{+0.09}_{-0.09}$ and $\sigma_{ext} = 0.42^{+0.03}_{-0.03}$ \cite{Huang2025h}, our analysis produces a somewhat lower slope with reduced statistical uncertainty.

\section{Conclusion and Discussion}
\label{conclusion}
In this work, we explore the use of Artificial Neural Networks (ANNs) and Bayesian Neural Networks (BNNs) for the calibration of observational Hubble data. Both approaches yield consistent reconstructions of the Hubble parameter, indicating that neural-network-based regression provides a stable and reliable framework for cosmological data calibration. From the reconstruction, we see that error bars increase with the paucity of data points for both ANN and BNN. This shows that the prediction in both the neural networks remain data-driven.\\

Despite this overall consistency, the Bayesian neural network framework offers clear conceptual and practical advantages. By explicitly incorporating prior information and treating the network weights probabilistically, the BNN naturally accounts for epistemic uncertainty arising from limited data and model flexibility. This results in more informative uncertainty estimates compared to the deterministic ANN approach, where uncertainties are typically inferred through repeated training or resampling techniques. Moreover, the large number of hyperparameters required in ANN architectures can make the resulting uncertainty estimates less robust. In addition, the Bayesian formulation provides improved control over model complexity and reduces the risk of overfitting, particularly in regimes where the data are sparse or noisy.\\

From a computational perspective, the ANN approach is comparatively faster and simpler to implement, making it well suited for exploratory analyses. However, for cosmological applications where a faithful propagation of uncertainties is essential, the BNN framework provides a more robust alternative. The ability of Bayesian Neural Networks to capture epistemic uncertainty makes them particularly valuable in cosmological studies, where accurate error propagation plays a crucial role in subsequent analyses. These results therefore indicate that the BNN approach is better suited for applications requiring reliable uncertainty quantification. This advantage becomes particularly important in the calibration of GRB correlations, where uncertainties in the reconstructed expansion history directly affect the inferred correlation parameters and their cosmological interpretation.\\

In the present work, we explore different model-independent neural-network reconstruction frameworks to address circulatory problem which remains a key issue in utilizing GRBs as cosmological probes. We restrict ourselves to Amati relation in order to study the impact of  different NN frameworks on model-independent GRB calibration and hence on Amati Relation Parameters. Nevertheless, several other GRB luminosity and energy correlations have been proposed in the literature. These correlations are significant for their potential cosmological applications. Extending the ANN and BNN calibration framework developed here to other GRB relations would provide an important insight into robustness of reconstruction methodology across different observables and correlation classes.\\

Several directions remain open for further investigation. Future studies may consider extending the present analysis to other GRB correlations in order to examine whether similar calibration behaviour arises across different observables.\\

Further work may also explore the inclusion of additional cosmological probes. The combination of GRB data with other independent datasets could potentially provide complementary information and help improve the robustness of the reconstruction and calibration procedure.\\

Bayesian Neural Networks may become increasingly valuable when analysing more complex cosmological datasets. As observational data grow in size and complexity, the ability of BNNs to provide principled uncertainty estimates and capture epistemic uncertainty will be particularly important. This feature makes them well suited for applications involving heterogeneous datasets or multi-probe cosmological analyses, where reliable uncertainty propagation is essential.\\

Finally, increasing availability of GRB observations from current and future missions is expected to significantly enlarge the high-redshift sample. This will further enable detailed comparisons between different calibration approaches, correlation relations, and redshift-dependent analyses. A larger and more homogeneous GRB dataset would allow for a more precise calibration of luminosity relations and enable GRBs to play a more prominent role as cosmological probes of the expansion history of the Universe.
\section*{Acknowledgments}
The authors thank the anonymous referee for valuable suggestions and comments which have helped us to enrich the manuscript. We would like to thank Dr. Tarun Kumar Gupta for his valuable guidance and discussions related to the Artificial Neural Network methodology employed in this study. N. Rani would like to thank IUCAA, Pune , India for providing her Vising Associateship under which a part of the work is carried out. 

\bibliographystyle{unsrtnat}
\bibliography{refs}

\end{document}